%% file: main.tex
\documentclass[lettersize,journal]{IEEEtran}
\usepackage{amsmath,amsfonts}
\usepackage[utf8]{inputenc}
\usepackage{array}
\usepackage[caption=false,font=normalsize,labelfont=normalfont,textfont=normalfont]{subfig}
\usepackage{url}
\usepackage{graphicx}
\usepackage{siunitx}
\usepackage{physics}
\usepackage{hyperref}
\usepackage{xcolor}
\usepackage[
    style=ieee,
    maxbibnames=2,
    minbibnames=1,
]{biblatex}
\usepackage{mathtools}

\newcommand{\s}{\text{s}}
\renewcommand{\i}{\text{i}}
\newcommand{\p}{\text{p}}
\newcommand{\inn}{\text{in}}
\newcommand{\out}{\text{out}}
\newcommand{\loss}{\text{loss}}
\newcommand{\add}{\text{add}}

\newcommand{\sys}{\text{sys}}
\newcommand{\exc}{\text{exc}}
\newcommand{\meas}{\text{meas}}

\newcommand{\E}{\mathcal E}

\makeatletter
\makeatother

\begin{document}

\title{
Device-Agnostic Microwave Noise Metrology for Nonlinear Cryogenic Quantum Devices}

\author{
Andrea~Celotto,
        Alessandro~Alocco,
        Bernardo~Galvano,
        Luca~Fasolo,
        Emanuele~Palumbo,
        Luca~Callegaro,
        Luca~Oberto,
        Patrizia Livreri,
        and~Emanuele~Enrico
\thanks{A. Celotto (andrea.celotto@polito.it), A. Alocco and E. Palumbo are with the Department of Applied Science and Technology of the Polytechnic University of Turin.}
\thanks{B. Galvano and P. Livreri are with the Department of Engineering of the University of Palermo.}
\thanks{A. Celotto, A. Alocco, B. Galvano, L. Fasolo, E. Palumbo, L. Callegaro, L. Oberto and E. Enrico are with the Istituto Nazionale di Ricerca Metrologica (INRiM).}%
}%

\maketitle

\begin{abstract}
Microwave devices capable of near-quantum-limited signal processing are essential components in the toolbox of solid-state quantum technologies. The manipulation and readout of single-photon microwave signals through amplifiers, mixers, isolators, etc. must fulfill strict requirements in terms of signal integrity to ensure reliable operation.

These active microwave quantum devices operate in complex cryo-electronic setups. This poses challenges to their characterization, since all relevant figures of merit must be expressed at the reference planes of their ports. Even though cryogenic S-parameter calibration is non-trivial, metrological approaches are converging toward rigorous methods. Furthermore, preserving signal integrity must be quantified via absolute noise levels at the ports of the Device Under Test (DUT), requiring an absolute power reference.

In this work, we present an \textit{in situ} noise metrology protocol based on substituting a controllable noise source for the DUT. We motivate this choice by showing that placing the noise source at the DUT input impacts the separability of the calibration from the DUT characteristics.

Our proposed architecture combines Planck spectroscopy using a Variable Temperature Stage with Short-Open-Load-Reciprocal scattering-parameter calibration, so that noise and scattering quantities are referred to the same cryogenic reference planes. In this configuration, the readout-chain calibration is separated from the internal dynamics of the DUT. As a demanding use case, we apply the protocol to a Josephson Traveling Wave Parametric Amplifier and extract its gain and input-referred added noise under pump conditions activating multimode nonlinear behavior. This illustrates how our device-agnostic protocol supports portable noise characterization of nonlinear cryogenic microwave devices.
\end{abstract}


\section{Introduction}\label{sec:intro}
\IEEEPARstart{C}{ryogenic} microwave quantum technologies rely on devices that process signals close to the single-photon level. Near-quantum-limited amplifiers are a prominent example because they enable high-fidelity readout of superconducting quantum circuits \cite{Aumentado_review}. At the same time, the microwave toolbox for quantum technologies includes frequency converters, mixers, magnetless isolators and circulators, nonlinear resonators, and other driven or active components \cite{Sliwa2015, Lecocq2017, Chapman2017, ranadive2025, malnou2025, Demarets2026}. These devices differ in function, but they share a common metrological difficulty: their noise must be measured in absolute units and referred to a well-defined plane inside a cryogenic measurement chain.

The quantity of interest depends on the DUT. For a phase-preserving amplifier, the relevant figure of merit is the input-referred added noise. For a frequency converter or mixer, one may instead need a conversion-added noise. For an isolator or a nonlinear resonator, a relevant observable may also be back-action noise. The general problem addressed here is the calibration of microwave noise at a cryogenic DUT reference plane, beyond the application of amplifier benchmarking.

For a phase-preserving active DUT with power transmission coefficient $G$, the 
output noise $N_\out$ includes the input fluctuations $N_\inn$ transmitted forward, and some added noise $N_\add$, introduced by the device itself. This quantity, which is referred to the device input, is not only due to internal dissipation, but there exists \cite{clerk} a fundamental limit on it, imposed by quantum mechanics:
\begin{equation}\label{eq:quantum_limit_on_noise}
    N_\add := \frac{N_\out}{G}-N_\inn \geq \frac{|G-1|}{2G}
\end{equation}
Here, noise is expressed as a photon flux: the photon number per unit time per unit bandwidth. This quantity is equivalent to the Power Spectral Density (PSD) divided by the photon energy $\hbar\omega$. In the high-gain limit $G \gg 1$, Eq.~\eqref{eq:quantum_limit_on_noise} gives the familiar half-photon quantum limit for a phase-preserving amplifier \cite{clerk}. 
Measuring the value of the added noise is essential to assess the suitability of a real device for quantum-limited signal processing at the single-photon regime.
Defining a systematic, self-contained, and simple experimental protocol for the measurement of this quantity is the goal of the present work.

\begin{figure}[]
    \centering
    \includegraphics[width=0.8\linewidth]{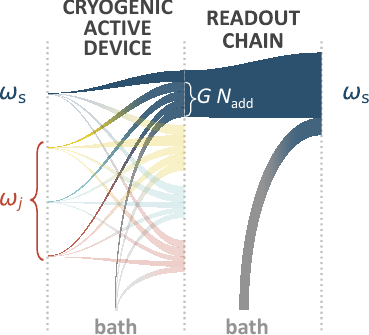}
    \caption{
    Flow diagram (not at scale) of the scattering of fluctuations impacting the output signal mode, tracked across relevant experimental stages. Photons are picked up from thermal baths due to dissipative processes. The gain $G$ can be identified as the ratio of the final and initial widths of the uppermost flow of photons inside the device. The same factor multiplies the added noise $N_\add$, as shown in the diagram.}
    \label{fig:noise_transfer}
\end{figure}

The way realistic active DUTs, depicted in Figure~\ref{fig:noise_transfer}, transmit forward input fluctuations is influenced by a large array of non-idealities, difficult to capture with analytical models.
These include spurious coupling to signal harmonics or intermodulation products, losses, and impedance mismatches. A purely behavioral representation of the input-referred added noise to an output signal at frequency $\omega_\s$ can be written as \cite{X_params_book, Floquet_TWPA, Peng_Xparameters, PTB_TWPA}
\begin{equation}\label{eq:actual_paramp_noise}
N_\add=
\sum_{\mathclap{(j, q)\neq(\s,1)}}
\frac{\left|S_{2 q}^{(\s j)}\right|^2}{G} N_{\inn, j}^{(q)}+ N_\loss,
\end{equation}
where $j$ labels the frequencies coupled to $\omega_\s$, $N_{\inn, j}^{(q)}$ are the input fluctuations at frequency $\omega_j$ and port $q$, and $S_{2 q}^{(\s j)}$ describes the scattering of these input fluctuations to the signal scattered out of port 2. The notation suggests that we may think of them as a multimode extension of the familiar S-parameters. Finally, $N_\loss$ accounts for loss-induced noise generated inside the DUT. Eq.~\eqref{eq:actual_paramp_noise} shows that the device-added noise depends on a variety of processes unknown \textit{a priori}. Therefore, if the extraction of absolute noise metrics relies on assuming a model for the device, distortions may be introduced, as we show in this work.

Additionally, Eq.~\eqref{eq:actual_paramp_noise} highlights that the added noise depends on the electronic temperature at its ports through $N_{\inn,j}^{(q)}$. Therefore, in order to operate a device as close as possible to the quantum limit, these input fluctuations should be reduced to the vacuum level, which in the microwave range corresponds to electronic temperatures as low as $\sim 10^{-2}$ K \cite{100qubits, Ranzani_calibration}.

Measuring absolute noise in a cryogenic environment is challenging, as it requires a robust absolute power reference to calibrate room-temperature measurements directly to the reference planes at the ports of the Device Under Test (DUT). Existing methods often place a calibrated noise source at the DUT input \cite{Malnou2021, Martinis2015, Malnou2024, RanzaniKITWPA, Ranadive2022, PTB_TWPA}. This serial arrangement is useful and widely adopted. However, for nonlinear or parametrically driven DUTs, the extraction of the DUT contribution can depend on the internal noise-transmission model assumed during calibration. A representative bias mechanism is discussed in Appendix~\ref{app:biases}.

Conversely, experimental configurations with the noise source substituting the DUT have been employed in the literature \cite{Zobrist, HoEom2012, klimovich2023}. These works succeeded in providing
an unbiased calibration of the readout chain, separating it from the behavior of the DUT. This is not sufficient by itself. The calibrated noise spectrum must still be combined with scattering information referred to the same microwave planes. Otherwise, heuristic gain estimates, for example, pump-on/pump-off comparisons in parametric amplifiers, can introduce systematic errors if impedance mismatches or losses are neglected.

Here, we present a noise characterization protocol that combines scattering parameters and a calibration of the readout chain performed independently from the DUT. This provides a comprehensive and general protocol for the characterization of noise in microwave quantum devices that can be readily adapted to many device architectures.

In our implementation, a Variable Temperature Stage (VTS) provides an \textit{in situ} thermal noise, that can be substituted to the DUT within the measurement network thanks to a pair of microwave switches. The same cryogenic switch network also supports Short-Open-Load-Reciprocal (SOLR) S-parameter calibration \cite{Ferrero1992, oberto2026}. We demonstrate the method on a Josephson Traveling Wave Parametric Amplifier (JTWPA), where the extracted observable is the added noise. This amplifier use case is deliberately demanding because multimode nonlinear processes and pump-dependent behavior can strongly affect the measured noise. 

\section{Theory of Cryogenic Noise Measurements}\label{sec:theory}

\subsection{The need for cryogenic microwave calibration}
Quantum devices operating in the microwave range require cryogenic temperatures as low as millikelvin. Therefore, they have to be placed inside complicated cryo-electronic apparatuses \cite{Ranzani_calibration, 100qubits}, whose impact on the measurement must be de-embedded through dedicated calibration protocols.

We distinguish between two types of calibration, regarding scattering parameters and absolute noise levels. The latter requires an absolute power reference to characterize the readout chain in terms of its gain and added noise, $G_\sys$ and $T_\sys$. 

\begin{figure}[]
    \centering
    \includegraphics[width=\linewidth]{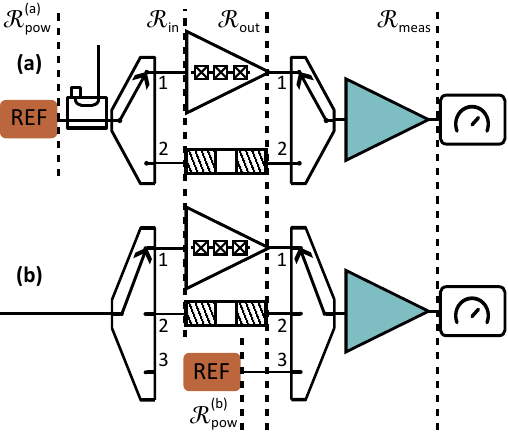}
    \caption{Two placements of the absolute noise reference, labeled REF. (a) In a serial configuration the DUT is included in the path between the noise source and the detector. (b) In a substitution configuration the noise source and the DUT are connected to the readout chain through cryogenic switches. The reference planes $\mathcal R_j$ indicate where the relevant microwave quantities are defined. The female-female \textit{thru} connection can symbolize a more complete S-parameter calibration calkit consisting also of impedance standards.}
    \label{fig:noise_meas_comparison}
\end{figure}

Assuming a linear readout chain, the measured PSD at frequency $\omega$ can be written as
\begin{equation}\label{eq:readout_chain}
    S_\meas(\omega)=G_\sys(\omega)\left[\hbar\omega N_\out(\omega)+k_BT_\sys(\omega)\right] ,
\end{equation}
The noise reference plane $\mathcal R_\out$, see Figure~\ref{fig:noise_meas_comparison}, at which $N_\out$ is defined, operationally sets where the "readout chain" begins. Acquiring $S_\meas$ for at least two known data points of $N_\out$ allows to retrieve both $G_\sys$ and $T_\sys$. 
Once they are known, any measured PSD can be converted into a noise photon flux at $\mathcal R_\out$. This absolute calibration is distinct from relative metrics such as signal-to-noise-ratio improvement, which are useful for benchmarking but do not by themselves establish an input- or output-referred noise photon number \cite{Planat, PTB_TWPA}.

Scattering-parameter calibration addresses a different question. It de-embeds the surrounding microwave network and returns relative transmission and reflection coefficients of the DUT. For an amplifier, this includes the gain $G$ needed in Eq.~\eqref{eq:quantum_limit_on_noise}. For other devices, it may instead provide reflection coefficients or multiport scattering parameters. Details about our implementation are described in Appendix~\ref{sec:exp_setup} and in \cite{oberto2026}. 

Noise calibration and scattering calibration are therefore complementary. A central point of the present work is that both must refer to the same cryogenic planes, and both calibrations must not be influenced by the behavior of the DUT, nor by any assumption about its physics.

\subsection{Predictable noise sources}\label{sec:noise_sources}
Absolute microwave-noise calibration can be pursued with cryogenic power sensors \cite{RF_transfer, Royal_Holloway} or with predictable noise sources. The latter are widely used in laboratories and literature, see \cite{Malnou2021, Martinis2015, Malnou2024, RanzaniKITWPA, Zobrist, HoEom2012, klimovich2023}, and they are the focus of this work. Three main approaches are used:

\paragraph{Hot-load / Y-factor}
Two matched loads at temperatures $T=T_\text{hot},T_\text{cold}$ are connected to a microwave switch \cite{Martinis2015, klimovich2023, HoEom2012} and emit thermal noise
\begin{equation}\label{eq:thermal_noise}
    N_\text{therm}(\omega;T) = \frac{1}{2} \coth\left(\frac{\hbar \omega}{2 k_B T}\right).
\end{equation}
The ratio between the two measured noise spectra, known as \textit{Y-factor}, gives the readout-chain gain $G_\sys$ and noise temperature $T_\sys$. This approach is conceptually simple, but it is sensitive to asymmetries between the two loads and the paths connecting them to the switch.

\paragraph{Variable Temperature Stage (VTS) / Planck spectroscopy}
A VTS provides a single thermal source whose temperature can be swept continuously \cite{Ranadive2022, Malnou2024, 2D_Planck}, a process known as Planck Spectroscopy. Fitting the measured PSD to Eq.~\eqref{eq:thermal_noise} over several temperatures improves the robustness against statistical fluctuation. A single continuously tunable source also avoids the asymmetries that the hot-load method is sensitive to. The VTS must be thermally decoupled from its environment while maintaining internal thermal equilibrium between its thermometer, heater, and microwave dissipative element.

\paragraph{Shot Noise Tunnel Junctions}
Shot Noise Tunnel Junctions provide a voltage-controlled emission of shot noise \cite{SNTJ_Spietz, SNTJ_Aumentado, Malnou2021, Malnou2024}. Sweeping a voltage is fast, therefore their use allows fast calibration. Their implementation is complicated by poor 50 $\Omega$ matching \cite{Malnou2024}, the need for bias-tees which introduce additional loss, and high sensitivity to electrostatic discharges.

\vspace{12pt}
Regardless of the particular noise source, special attention must be placed on its non-ideality, mainly concerning impedance matching. We generalize a noise source as a 2-port component emitting noise out of port 2:
\begin{equation}\label{eq:actual_noise_source_output}
    N_\text{pow}=(1-|S_{22}|^2)N_\text{source} + \frac{1}{2}|S_{21}|^2,
\end{equation}
assuming vacuum fluctuations entering port 1. Here $N_\text{source}$ may be the thermal noise of Eq.~\eqref{eq:thermal_noise} or shot noise. As long as $|S_{21}|^2 \ll 1-|S_{22}|^2$, any matched 2-port component, or a 1-port component whose port is labeled with 2, can be treated as a practical noise source. This validates the implementation of a 20 dB attenuator as the thermal noise source used in this work.

\section{Serial and Substitution Placement of the Noise Reference}\label{sec:absolute_cal}

As previously mentioned, in order to characterize the noise added by any active device, both noise and scattering quantities need to be defined at precise reference planes coinciding with the DUT ports, labeled as $\mathcal{R}_\inn$ and $\mathcal{R}_\out$ in Figure~\ref{fig:noise_meas_comparison}. Two configurations are considered here, differing by whether the noise emitted by the calibrated source is propagated through the DUT during readout-chain calibration.

\subsection{Serial configuration, applied to a parametric amplifier}
In a serial configuration, see Figure~\ref{fig:noise_meas_comparison}a, the noise source is placed at the input of the DUT. This forces the known noise radiation to travel from the source output plane $\mathcal R_\mathrm{pow}^{(a)}$ to RT instrumentation at plane $\mathcal R_\meas$ only through the DUT. 

This means that, in this configuration, the readout chain to be calibrated includes the DUT itself. If the DUT in question is not linear, then \eqref{eq:readout_chain} does not hold anymore, and a model for the propagation of noise through the DUT must be assumed, since the multimode scattering amplitudes $|S_{2q}^{(\s j)}|^2$ are not known \textit{a priori}. This "forces" us to consider a particular device architecture to further analyze the serial configuration.

In particular, here we consider a parametric amplifier \cite{Caves_Schumaker}, which works as a nonlinear medium that amplifies a signal at frequency $\omega_\s$ when pumped with a strong tone, called the \textit{pump}, at frequency $\omega_\p$. Physically, parametric amplification occurs through a wave-mixing process: a photon from the pump converts into a photon at the signal frequency and one at the idler frequency $\omega_i$. For an ideal, large gain, parametric amplifier \cite{clerk}, the output noise on the signal is
\begin{equation}\label{eq:ideal_paramp}
    N_{\out, \s}= G ( N_{\inn, \s} + N_{\inn, \i}),
\end{equation}
which means that the added noise consists of the input fluctuations on the idler, thus achieving quantum-limited amplification if those are vacuum, i.e. if $N_\add \equiv N_{\inn, \i}=\frac{1}{2}$.

For a real device, the added noise exceeding the quantum limit is introduced to \eqref{eq:ideal_paramp}, reading
\begin{equation}\label{eq:real_paramp}
    N_{\out, \s}= G ( N_{\inn, \s} + N_{\inn, \i}+N_\exc).
\end{equation}
Note that the added noise is minimized for input vacuum fluctuations, i.e. for $N_\add = N_{\inn, \i}+N_\exc=\frac{1}{2}+N_\exc$.

We point out that this model, as explained in \cite{Malnou2021, Malnou2024}, has the significant advantage of accounting for input idler fluctuations, something that many, especially in the early works showcasing noise characterization of superconducting parametric amplifiers, failed to do. This meant a generous underestimation of the added noise $N_\add$, by roughly a factor of 2.

A model for the serial calibration method for the measured PSD at room temperature can be written as
\begin{subequations}\label{eq:line_noise_model_a}
\begin{align}
S_\meas &= \hbar \omega_\s \tilde G_\sys \left(N_{\text{pow},\s}+\frac{A_\i}{A_\s}N_{\text{pow},\i}+\tilde N_\sys\right) ,\label{eq:serial_meas}\\
\tilde G_\sys &:=G_\sys G A_\s ,\label{eq:serial_gain}\\
\tilde N_\sys &:=\frac{2-A_\s-A_\i}{2A_\s}+\frac{N_\exc}{A_\s}+\frac{k_BT_\sys}{\hbar\omega_\s G A_\s} .\label{eq:serial_noise}
\end{align}
\end{subequations}
Here $A_\s$ and $A_\i$ describe losses between the noise source and DUT at the signal and idler frequencies, respectively, while $N_\exc$ collects excess contributions beyond the ideal two-mode model. The quantities $\tilde G_\sys, \tilde N_\sys$ can be obtained by regressing $S_\meas$ against $N_\text{cal}:=N_\text{pow,s}  +  (A_\i/A_\s) N_\text{pow,i}$.

This model relies on the assumption of constant $N_\exc$, which is fundamentally flawed. As \eqref{eq:actual_paramp_noise}, $N_\exc$ cannot be a constant, since it must account not only on the dissipation-induced contribution $N_\loss$, but also on additional input fluctuations scattered to the signal. This makes the noise at the DUT output dependent on the noise level emitted by the noise source at its input, ultimately distorting the estimates on $\tilde G_\sys$ and $N_\exc$ itself. Appendix~\ref{app:biases} gives an analytical treatment of this effect, and gives a quantitative estimation of the underestimation of noise exploiting synthetic but realistic data. Our conclusion is not that serial configurations are intrinsically invalid, but that their interpretation can become model-dependent for nonlinear devices with unmodeled noise channels.

Some additional remarks must be made on the other parameters present in the model. The attenuations $A_\s, A_\i$  must be separately characterized, as in \cite{Malnou2024}, or taken from datasheets, which are notoriously lacking in terms of mK characterization. The gain $G$ can be estimated either by comparison of the transmission to that of the \textit{thru} reference, or by implementing a full S-parameter calibration protocol, naturally enabled by placing additional standards on the other ports of the switches. In addition, the components between $\mathcal R_\out$ and $\mathcal R_\meas$ can be characterized in terms of $G_\sys$ and $T_\sys$ by excluding the DUT using the switches, and then repeating the calibration.

Finally, we must point out that the noise at $\mathcal R_\mathrm{pow}$ may not coincide with the prediction of thermal or shot noise laws, because of impedance mismatches. Therefore, the S-parameters of the noise source should be separately measured in a dedicated cooldown. This is needed to predict $N_\mathrm{pow}$ from the source's physical temperature or bias voltage, as \eqref{eq:actual_noise_source_output} shows.

\subsection{Substitution configuration}
In a substitution configuration, the noise source is connected to the same readout chain as the DUT, but through a different switch throw. This mitigates nonlinear distortion because the readout chain does not include the active DUT, thereby satisfying the assumption of overall system linearity as in Eq. \eqref{eq:readout_chain}. The RT measurement of the noise PSD emitted by the noise source is
\begin{subequations}\label{eq:line_noise_model_b}
\begin{align}
S_{\meas} &= \hbar\omega_\s \tilde G_{\sys}\left(N_{\text{pow},\s}+\tilde N_{\sys}\right),\label{eq:parallel_psd}\\
\tilde G_{\sys} &= G_{\sys}\,A_\s,\label{eq:parallel_gain}\\
\tilde N_{\sys} &= \frac{1-A_\s}{2A_\s}+\frac{k_BT_{\sys}}{\hbar\omega_\s A_\s} .\label{eq:parallel_noise}
\end{align}
\end{subequations}
A simple linear least squares regression yields $G_{\sys}$ and $T_{\sys}$, provided $A_\s$ is known. After calibration, switching back to the DUT gives
\begin{subequations}\label{eq:dut_output_noise}
\begin{align}
S^{(\text{DUT})}_{\meas} &= G_{\sys}\left(\hbar\omega_\s N_{\out,\s}+k_BT_{\sys}\right),\label{eq:dut_meas}\\
N_{\out, \s} &= G\left(N_{\inn,\s}+N_{\add}\right) .\label{eq:dut_noise}
\end{align}
\end{subequations}
Eq.~\eqref{eq:dut_noise} is the amplifier-specific interpretation used in the demonstration below. For another DUT, Eq.~\eqref{eq:dut_meas} would still provide the calibrated output spectrum, but the final observable might not be the added noise $N_\add$ but a more telling noise metric defined in terms of input- or output-referred noise and calibrated S-parameters.

Here, the device transmission $G$ must be estimated either by comparison with a \textit{thru} reference, or with a full S-parameter calibration, naturally enabled by the switches, as for the serial configuration.

Note that combining the gain $G$ extracted from a coherent transmission measurement with the noise power $N_{\out}$ from the amplification of pure noise relies on a strict small-signal assumption. We assume that the device's internal wave-mixing dynamics, transmission coefficients, and added-noise mechanisms remain identical whether the device is probed by a weak coherent tone or is only processing ambient input fluctuations.

The substitution configuration can naturally solve the problems of the serial one. First, the calibration of the readout chain is completely decoupled from the characterization of the DUT, making the method robust to the behavior of any practical DUT. Secondly, the attenuation $A_\s$ can be neglected if a VTS is chosen as the noise source, since it can be connected to the switch directly with a superconducting cable, for which $A_\s\approx1$, making $\mathcal{R}_\text{pow}^\text{(b)}$ and $\mathcal{R_\out}$ effectively coincide. Furthermore, placing the noise source between the switches naturally allows for \textit{in-situ} calibrated measurement of its S-parameters.

Finally, in order to extract $N_\add$, the signal input noise $N_{\inn, \s}$ must be known. A proper distribution of attenuation along the input lines, see \cite{100qubits} and Section~\ref{sec:setup} in \cite{SupplMat}, can ensure they correspond to vacuum fluctuations.

\section{Experimental Implementation}\label{sec:exp_setup}

\subsection{Cryoelectronic setup and VTS}
The experimental platform is designed for cryogenic two-port characterization of nonlinear microwave devices in the C and X bands. 
Input lines are attenuated and thermalized at several temperature stages, while output signals are routed through cryogenic isolators, HEMT amplification at 3 K, and room-temperature amplification before detection by a Nonlinear Vector Network Analyzer (NVNA). Further setup details are reported in the Supplementary Materials \cite{SupplMat}.

The DUT and the VTS are mounted at the coldest stage of a dilution refrigerator between a pair of cryogenic electromechanical RF switches.
Our VTS consists of a matched $-20$ dB attenuator mounted on a thermally isolated copper stage equipped with a heater and thermometer. Details on its particular realization are provided in Appendix~\ref{app:VTS}. 

Mounting both the DUT and the VTS between switches naturally gives the opportunity to equip the switches with impedance and line standards in order to perform a full S-parameter calibration protocol.
Different calibration protocols have been implemented in cryogenic environments \cite{Ranzani_calibration, NPLcal}. Here we use the Short-Open-Load-Reciprocal (SOLR) technique \cite{Ferrero1992}, which essentially uses Short, Open and Load impedance standards for separate single-port calibrations. Additionally, any reciprocal two-port network can serve as the reference needed to obtain the full transmission tracking of the measurement apparatus. The VTS attenuator used in this work therefore can also act as a reciprocal calibration element. Our cryogenic implementation of SOLR, together with a study on its traceability and uncertainty budget, is discussed in Ref.~\cite{oberto2026}.

\subsection{Workflow}\label{sec:workflow}
We propose a standardized workflow for characterizing the gain and added noise of a microwave cryogenic quantum device independently from any assumed physical model for the device itself:
\begin{enumerate}
    \item connect the DUT, VTS, and scattering standards between a pair of cryogenic microwave switches
    \item perform SOLR calibration
    \item measure the DUT scattering parameters at the selected operating point, ensuring a linear operating regime
    \item measure the noise PSD of the device in the selected working point (without seeding it with a signal)
    \item measure the VTS S-parameters to verify that it behaves as a suitable thermal source for the calibrated readout port;
    \item sweep the VTS temperature and perform Planck spectroscopy to extract $G_\sys(\omega)$ and $T_\sys(\omega)$ of the readout chain;
    \item switch back to the DUT, measure its output PSD at the chosen operating point, and convert the result to $N_\out(\omega)$ using Eq.~\eqref{eq:readout_chain};
    \item combine the calibrated output spectrum with the device-specific scattering data to extract the desired observable.
\end{enumerate}

For the JTWPA use case, the relevant observable is the input-referred added noise. The NVNA allows both the scattering measurement and the spectrum-analyzer functionality used to record PSDs. Its dual functionality has the great benefit of yielding a single reference plane for all RT measurements.

Crucially, combining the scattering measurement (Step 3) from the noise measurement (Step 4) requires the DUT to operate in a perfectly linear regime with respect to the signal mode. The coherent probe tone applied in Step 3 must be sufficiently weak so that it does not induce pump depletion, alter the operating point, or activate nonlinearities that are not present during the unseeded noise measurement in Step 4.

The JTWPA used here is a pre-commercial AI-TWPA-C from Arctic Instruments \cite{JTWPA_patent, VTT_TWPA_paper}, based on a series array of flux-tunable Superconducting Nonlinear Asymmetric Inductive eLements (SNAILs) \cite{SNAIL}. A DC flux bias tunes the three-wave-mixing response. The pump was intentionally chosen so that the device operates outside an ideal low-noise regime and exhibits multimode behavior. To ensure the small-signal assumption holds between the gain and noise measurements, the coherent VNA probe tone was kept heavily attenuated (e.g., at a nominal power of $\sim -120$ dBm at the device input), ensuring it did not induce pump depletion or alter the nonlinear operating point compared to the unseeded state. This operating point is not meant to optimize amplifier performance. It is used to test whether the calibrated readout protocol can track strong pump-dependent changes in output noise without embedding a simplified DUT model into the calibration itself.

\section{Use Case: Added Noise of a JTWPA}\label{sec:results}

\begin{figure*}
    \centering
    \includegraphics[width=0.8\linewidth]{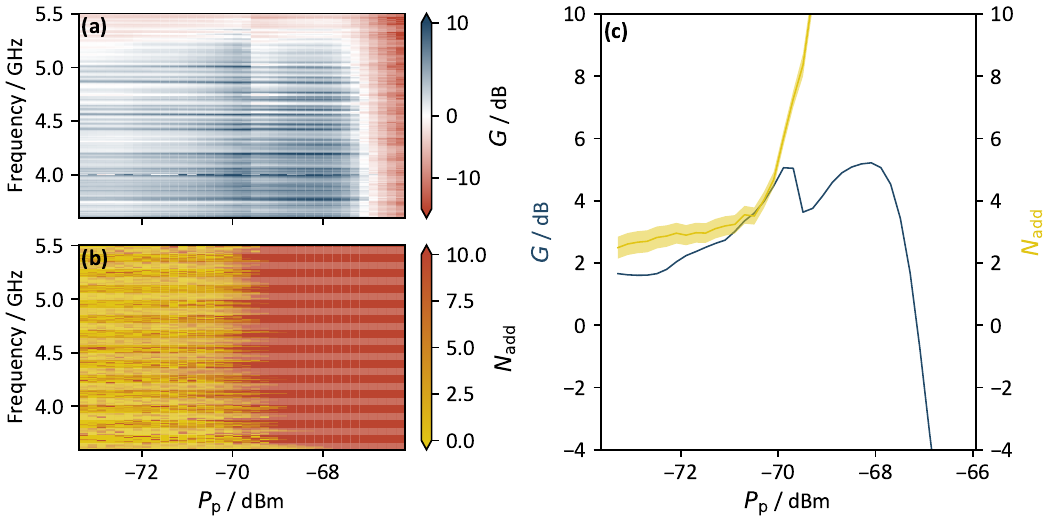}
    \caption{(a,b) Gain $G$ and added noise $N_\add$ of the JTWPA as functions of pump power $P_\p$ and signal frequency. The frequency-dependent fine features in the gain, commonly referred to as ripple, are deterministic features attributed to standing waves in the device packaging rather than measurement noise, as their consistency for various pump powers may suggest. (c) Gain and added noise, averaged over the 3.6--5.5 GHz band, as functions of pump power. The colored band around the noise curve represents its associated uncertainty, propagated from the fit on Planck-spectroscopy data.}
    \label{fig:Nadd}
\end{figure*}

Figure~\ref{fig:Nadd} shows the measured JTWPA gain and input-referred added noise as functions of pump power and signal frequency. Because the device is deliberately operated under non-optimal pump conditions, the performance should not be interpreted as a benchmark of the best achievable JTWPA noise. Instead, the measurement is used as a stress test for the calibration protocol.

At low pump powers, the band-averaged gain and added noise increase together. At higher pump powers, the added noise grows much more rapidly than the gain. This behavior is consistent with the activation of additional pump-dependent noise channels, such as higher-order wave-mixing processes or pump-enhanced loss. Such channels are expected in realistic nonlinear traveling-wave devices and have been discussed as limiting mechanisms for quantum-limited amplification \cite{Floquet_TWPA, Peng_Xparameters}.

The key point for the present work is not the microscopic identification of the dominant excess-noise mechanism. Rather, it is that the output noise is measured after the readout chain has been calibrated independently of the JTWPA internal scattering model. The same data can therefore be used as input for more detailed physical modeling, but the extraction of the calibrated output spectrum does not require assuming a two-mode amplifier model during the readout-chain calibration.

\section{Conclusions and Perspectives}\label{sec:conclusions}
We have presented an \textit{in situ} microwave noise-metrology protocol for cryogenic quantum devices. The protocol combines a VTS thermal noise reference, to be substituted to the DUT exploiting microwave switches, with SOLR scattering-parameter calibration. This architecture aligns absolute-noise and scattering measurements at the same cryogenic reference planes and separates readout-chain calibration from the nonlinear dynamics of the device under test.

The method is general at the level of calibrated noise referencing. 
The protocol does not remove the need to define the relevant observable and operating point of the DUT. Instead, it provides a calibrated reference-plane framework in which such observables can be measured more reproducibly.

As a use case, we applied the protocol to a JTWPA and extracted gain and input-referred added noise under pump conditions that activate multimode nonlinear behavior. The results show a pump-dependent increase of added noise that would be difficult to interpret reliably if the readout-chain calibration itself relied on a simplified internal model of the amplifier. More broadly, the approach provides a practical route toward standardized noise characterization of nonlinear cryogenic microwave devices used in quantum technologies.

The present implementation is not a primary noise standard. Its accuracy depends on the VTS thermal model, impedance characterization, reference-plane calibration, readout-chain linearity, and the uncertainty budget of the Planck-spectroscopy fit. A complete quantitative analysis of VTS thermal gradients and traceability of its implementation is therefore an important next step. Once these uncertainty contributions are fully assessed, the same architecture can support automated benchmarking of multiple cryogenic microwave devices within a single dilution-refrigerator platform.
\let\normalsection\section
\appendices

\section{Analytical Example of Bias in a Serial Configuration}\label{app:biases}
This appendix summarizes one mechanism by which a simplified serial noise calibration can become biased when the DUT has unmodeled energy-dependent noise channels. The derivation is not a proof that every serial method fails. It identifies a specific bias that can arise when a nonlinear DUT is included in the calibration path and the data are fitted with an insufficient noise-transmission model.

To treat thermal and shot-noise sources in the same notation, introduce a dimensionless control parameter
\begin{equation}
    \E := \frac{k_B T}{\hbar\omega_\s} \quad \text{or} \quad \E := \frac{eV}{2\hbar\omega_\s},
\end{equation}
so that the source noise can be represented as a function $N_\text{source}(\omega,\E)$ that is monotonic and convex with respect to the relevant experimental knob, captured by $\E$. In both cases, the asymptotic limit for large $|\mathcal E|$ is $\hbar \omega_s |\mathcal E|$.

The measured noise in a serial configuration can be decomposed as
\begin{subequations}\label{eq:decomposed_model}
    \begin{gather}
        N_\text{meas}(\E) = \tilde G_\text{sys} \left[N_\text{cal}(\E)+\tilde N_\text{sys}(\E)\right], \label{eq:full_nonlinear_readout_noise}\\
        N_\text{meas}(\E) := S_\text{meas}(\E) / \hbar \omega_\text{s},\\
        N_\text{cal}(\E):= N_\text{source}(\omega_\text{s}, \E)+ \frac{A_\text{i}}{A_\text{s}} N_\text{source}(\omega_\text{i}, \E), \\
        \tilde N_\text{sys}(\E)=N_0+\delta N (\E),\\
        \delta N (\E):=\sum_{\substack{n \in \mathbb Z \\ n \neq 0,-1}} 
        \frac{A_n |S_{2 1}^{(\s n)}|^2}{ A_\text{s} G}
        N_\text{source}(|\omega_\text{s}+n\omega_\text{p}|, \E),\\
        \begin{aligned}
            N_0 =& \frac{2-A_\text{s}-A_\text{i}}{2A_\text{s}}+\frac{k_B T_\text{sys}}{\hbar \omega_\text{s} G A_\text{s}} \\
            &+ \frac{1}{GA_\text{s}}\left[\frac{1}{2}\sum_{n\in\mathbb Z}|S_{2 2}^{(\s n)}|^2+
            \sum_{\substack{n \in \mathbb Z \\ n \neq 0,-1}} \frac{1-A_n}{2}|S_{2 1}^{(\s  n)}|^2\right].
        \end{aligned}
    \end{gather}
\end{subequations}
Here $A_n=A(|\omega_\text{s}+n\omega_\text{p}|)$ and $S_{2 i}^{(\s n)}$ represent scattering amplitudes to the output signal.

If Eq.~\eqref{eq:full_nonlinear_readout_noise} is fitted with the simplified affine model $N_\text{meas}=\tilde G_\text{sys}(N_\text{cal}+\tilde N_\text{sys})$, the dependence of $\tilde N_\text{sys}$ on $\E$ is neglected. Denoting averages over sampled values of $\E$ by $\langle\cdot\rangle$, the least-squares distortion is governed by
\begin{equation}
    \beta:=\frac{\mathrm{Cov}(\delta N,N_\text{cal})}{\mathrm{Var}(N_\text{cal})}.
\end{equation}
The fitted quantities become
\begin{subequations}\label{eq:bias_estimators}
\begin{align}
\tilde G_\text{sys,fit} &= \tilde G_\text{sys}(1+\beta),\\
\tilde N_\text{sys,fit} &= \frac{N_0 + \langle \delta N \rangle - \beta\langle N_\text{cal} \rangle}{1+\beta}.
\end{align}
\end{subequations}
For the representative case considered here, $N_\text{cal}(\E)$ and $\delta N(\E)$ increase together, giving $\beta>0$ and therefore an overestimate of the effective system gain.

The corresponding error on the fitted noise is
\begin{equation}\label{eq:bias_noise}
\tilde N_\text{sys,fit}-\tilde N_\text{sys}(0)
= \frac{\left[\langle \delta N \rangle -\delta N(0)\right] - \beta\left[\langle N_\text{cal} \rangle +\tilde N_\text{sys}(0)\right]}{1+\beta}.
\end{equation}
When the calibration samples the high-energy regime, the statistical moments are dominated by their asymptotic behavior:
\begin{equation*}
    \begin{aligned}
        \langle N_\text{cal}\rangle  \simeq \alpha\langle|\E|\rangle, \qquad
        \langle\delta N\rangle \simeq \gamma \langle|\E|\rangle, \qquad
        \beta \simeq \gamma / \alpha,
    \end{aligned}
\end{equation*}
with
\begin{equation*}
\alpha := 1+\frac{A_\text{s}\omega_\text{s}}{A_\text{i}\omega_\text{i}}, \qquad
\gamma := \sum_{n \neq 0,-1} \frac{A_n |S_{2 1}^{(\s n)}|^2} \omega_\text{s}{A_\text{s}G|\omega_\text{s}+n\omega_\text{p}|}.
\end{equation*}
In this limit,
\begin{equation}\label{eq:bias_final}
\tilde N_\text{sys,fit}-\tilde N_\text{sys}(0)\simeq 
-\left[\delta N(0)+\frac{\beta}{1+\beta}N_0\right]<0.
\end{equation}
Under the stated assumptions, the simplified serial fit underestimates the zero-temperature effective noise contribution. Figure~\ref{fig:distorted_fit} illustrates the effect using synthetic data generated as described in the Supplementary Materials \cite{SupplMat}. Although the quantification of the bias relies on synthetic data, they are generated starting from experimental quantities

\begin{figure}[t]
    \centering
    \includegraphics[width=\linewidth]{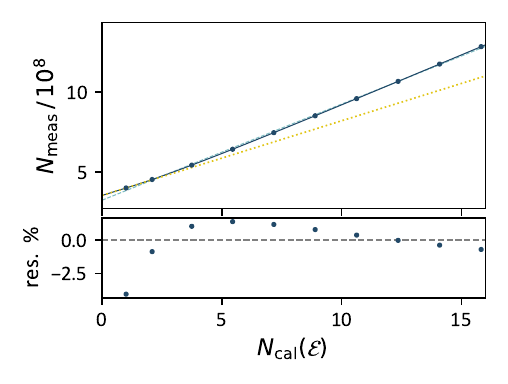}
    \caption{Illustration of the bias described in Appendix~\ref{app:biases} based on synthetic but realistic data. The solid curve is the nonlinear model of Eq.~\eqref{eq:full_nonlinear_readout_noise}; the dots are sampled calibration points; the dashed line is the simplified affine regression. The clear structure of residuals to the fit shows the consequence of neglecting energy-dependent contributions $\delta N(\E)$.}
    \label{fig:distorted_fit}
\end{figure}

\section{Technical Details of the VTS}\label{app:VTS}
\begin{figure}[t]
    \centering
    \includegraphics[width=0.5\linewidth]{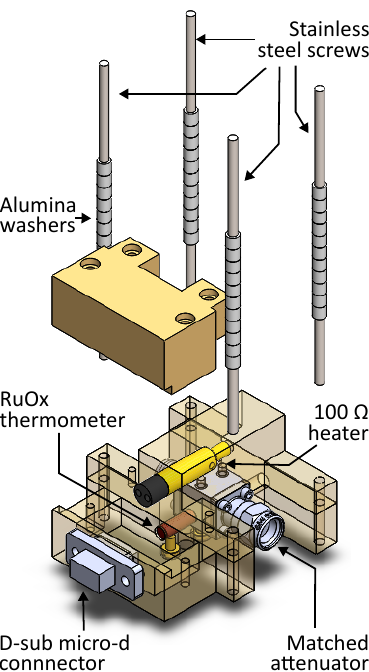}
    \caption{Mechanical rendering of the custom VTS assembly. The internal cavity houses a matched attenuator, a \SI{100}{\ohm} heater, and a ruthenium-oxide thermometer. The VTS is mounted on the sample space from below, anchored through stainless-steel screws. Alumina washers are used to control the physical distance of the object from the samples mounted above.}
    \label{fig:VTS_render}
\end{figure}

A rendering of the VTS assembly is shown in Figure~\ref{fig:VTS_render}. The main body is gold-plated copper, chosen to improve thermal contact at mechanical interfaces. It contains a 20 dB attenuator, a \SI{100}{\ohm} heating resistor, and a ruthenium-oxide thermistor (LakeShore RX-102A-AA). The heater and thermometer are biased and read out in a four-terminal configuration; the thermistor is measured with a LakeShore 372 AC resistance bridge.

The VTS is mechanically connected to the sample space through stainless-steel M2 screws and separated from the samples with alumina washers. In the present implementation its resting temperature is approximately 120 mK. Heating up to 2 K was obtained with powers in the 10 nW range, well below the cooling power of the dilution refrigerator mixing chamber.

Both DC and RF connections use superconducting NbTiN wiring. Because superconductors well below their critical temperature are poor thermal conductors, heat leaks through the coaxial conductors are strongly suppressed. The dominant thermal link is therefore expected to be the macroscopic mechanical connection. The electron temperature of the resistive microwave element is assumed to follow the temperature of the gold-plated copper body measured by the thermistor. The present thermal design is sufficient for the methodological demonstration reported here. A complete quantitative analysis of thermal gradients inside the VTS, needed for primary metrological traceability, is the subject of ongoing work.

\section*{Acknowledgments}
We would like to thank the team at VTT Technical Research Centre of Finland Ltd for providing the JTWPA: Debopam Datta, Wisa Forbom, Joonas Govenius, Robab Najafi Jabdaraghi, Janne Lehtinen, Jaani Nissila, Mika Prunnila, Jorden Senior, Nils Tiencken, and Visa Vesterinen.\\
This work is supported by the European projects Qu-Test, MiSS and MetSuperQ. 
Qu-Test has received funding from the European Union’s Horizon Europe – The EU research \& innovation programme under the Grant Agreement number 101113901. MiSS is funded by the European Union through the Horizon Europe 2021-2027 Framework Programme, Grant agreement ID: 101135868. The 23FUN08 MetSuperQ project has received funding from the European Partnership on Metrology, co-financed from the European Union's Horizon Europe Research and Innovation Programme and by the Participating States. 
Certain commercial equipment, instruments, or materials are identified in this paper to specify the experimental procedure adequately. Such identification is not intended to imply recommendation or endorsement by the authors, nor is it intended to imply that the materials or equipment identified are necessarily the best available for the purpose.

\printbibliography

\end{document}


\maketitle

\section{Cryoelectronic setup}\label{sec:setup}

\begin{figure}[h!]
    \centering
    \includegraphics[width=0.9\linewidth]{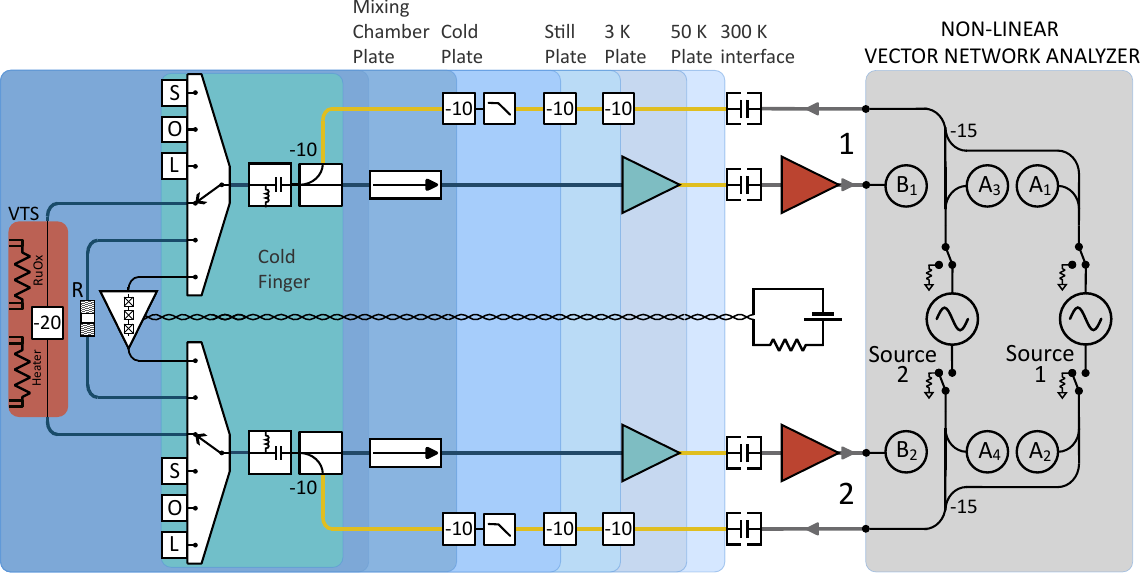}
    \caption{Cryo-electronic schematic of the measurement setup configured for the nonlinear characterization of the JTWPA. The diagram illustrates the attenuation and thermalization stages within the dilution refrigerator, scaling from room temperature (RT) down to the mixing chamber (MXC) at 20 mK. Key elements include the Device Under Test (DUT), the Variable Temperature Stage (VTS) used as an absolute noise reference, and the RF electromechanical switches employed for the Short-Open-Load-Reciprocal (SOLR) S-parameter calibration. Signal amplification is provided by a 3 K High Electron Mobility Transistor (HEMT) and RT amplifiers before detection by the Nonlinear Vector Network Analyzer (NVNA). This architecture ensures the decoupling of the absolute power reference from the nonlinear dynamics of the DUT.}
    \label{fig:setup}
\end{figure}

The cryo-electronic setup used at INRiM, shown in Fig. \ref{fig:setup}, is suited for two-port nonlinear characterization of TWPAs and other microwave active and passive devices operating in the C and X bands. The DUT is mounted at the coldest stage of a dilution cryostat (CF-CS110-600 from Leiden Cryogenics), and is connected to two electro-mechanical RF SP6T switches (R5927B2141), which are also connected to two sets of RF impedance standards (XMA 2001-7010-0X-CRYO) and the VTS. Bias tees (ZX85-12G-S+) are connected to the switch common ports, followed by directional couplers (QMC-CRYOCOUPLER-10) that couple (with -10 dB coupling factor) the input lines with the readout chains. Following the latter, isolators (LNF-ISISISC4\_12A) thermalized to the mixing-chamber plate prevent higher-temperature noise or reflections on subsequent stages from reaching the device. Amplification is provided by HEMT amplifiers (LNF-LNC0.3\_14A, light blue in Fig. \ref{fig:setup}) at 3 K and by room-temperature amplifiers (ZVA-183-S+, red in Fig. \ref{fig:setup}), while inner–outer DC blocks (BLK-18W-S+) mounted outside the cryostat decouple its ground from the RF connections. The amplified signals are sent to the receivers of a Nonlinear Vector Network Analyzer (NVNA) via SMA coaxial cables (SUCOFLEX 126E).

The NVNA (PNA-X N241B) features two microwave sources whose outputs are internally combined using the instrument's couplers: Source 1, with -15 dB coupling, generates the signal, while Source 2 generates the pump. The combined stimulus enters the cryostat through DC blocks, with -10 dB attenuators placed at the 3 K, still, and cold plates, where infrared filters are also located. RF wiring inside the cryostat uses both superconducting (C-SPSP-NBTI047-170MMNM, blue in Fig. \ref{fig:setup}) and cupronickel (SC-119/50-CN-CN) SMA coaxial cables (gold in Fig. \ref{fig:setup}). If required, a DC flux bias can be applied to the DUT via a shielded twisted pair connected to a floating voltage source (Yokogawa GS200), shunted by a 10 k$\Omega$ resistor to suppress current noise.

\begin{table}[]
\centering
\begin{tabular}{|l|r|r|r|r|r|r|}
\hline
\multicolumn{1}{|c|}{Stage} & \multicolumn{1}{c|}{$T$ / K} & \multicolumn{1}{c|}{$A_\text{lump}$ / dB} & \multicolumn{1}{c|}{$\ell$  / m} & \multicolumn{1}{c|}{$\alpha$ / (dB m$^{-1}$)} & \multicolumn{1}{c|}{$A_\text{tot}$ / dB} & \multicolumn{1}{c|}{$N$} \\ \hline
RT                          & 295{\color{white}.00}                        & 15                                     & 5                           & 0.85                                       & 19.25                                               & 1229.36                  \\
50 K                        & 50{\color{white}.00}                          & 0                                      & 0.25                            & 18.98                                        & 4.74                                                & 550.80                   \\
3 K                         & 2.55                         & 10                                     & 0.15                            & 18.98                                        & 12.85                                               & 38.70                    \\
Still                       & 0.82                         & 10                                     & 0.10                            & 18.98                                        & 11.90                                               & 5.72                     \\
CP                          & 0.13                         & 10                                     & 0.10                            & 18.98                                        & 11.90                                               & 1.02                     \\
MXC                         & 0.02                         & 0                                      & 0.15                            & 18.98                                        & 2.85                                                & 0.77                     \\
Cold Finger                          & 0.07                         & 10                                     & 0                             & -                                             & 10.00                                               & 0.56                     \\ \hline
\end{tabular}
\caption{Distribution of attenuation along the various thermal stages of the input lines. The $i$-th stage has temperature $T_i$, a lumped attenuation $A_{\text{lump},i}$ given by attenuators, some cables with length $\ell_i$ and attenuation per unit length $\alpha_i$. The total attenuation $A_{\text{tot},i}=A_{\text{lump},i} \cdot 10^{\alpha_i \ell_i / 10}$ allows to compute the noise photon flux transmitted to the next stage, $N_i$.}
\label{tab:input_lines}
\end{table}

Taking $\omega = 2\pi\,\cdot\,\SI{5}{GHz}$ as a reference frequency, the input lines provide 73.48 dB of nominal attenuation, whose distribution between the thermal stages ensures as low as 0.56 input thermal noise photons, as shown in Table \ref{tab:input_lines}. This number is found by keeping track of the physical temperatures $T_i$ of the various thermal stages, starting from the room temperature $T_0$=\SI{295}{K}, together with the total attenuation $A_{\text{tot},i}$ introduced by components at each stage, and then recursively applying \cite{100qubits}
\begin{gather*}
    N_0(\omega)=N_\text{therm}(\omega, T_0), \quad  \text{with} \quad
    N_\text{therm}(\omega, T)=\frac{1}{2}\coth\left(\frac{\hbar\omega}{2k_BT}\right)\\
    N_i(\omega)=A_{\text{tot},i}(\omega) N_{i-1}(\omega)+
    \big[1-A_{\text{tot},i}(\omega)\big] \,N_\text{therm}(\omega,T_i), \quad \text{for} \quad    i > 0
\end{gather*}

\section{Preliminary device characterization}\label{sec:pre_charac}

\subsection{Definition of measurement frequency range}\label{sec:freq_range}

\begin{figure}[h]
    \centering
    \includegraphics{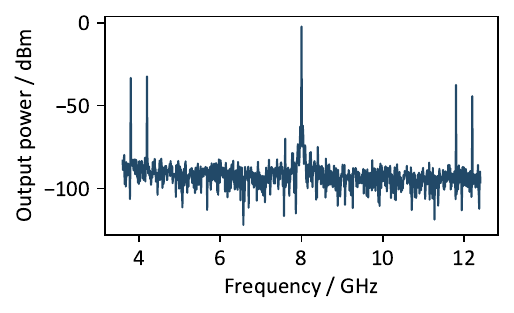}
    \caption{Spectrum showing JTWPA response under pumping at 8 GHz and probing at 3.8 GHz. The power level was scaled to the JTWPA output port with the system gain extracted with the Planck Spectroscopy.}
    \label{fig:freq_comb}
\end{figure}

Figure \ref{fig:freq_comb} illustrates the resulting spectral content generated when pumping the device at $\omega_\text{p} = 2\pi \cdot \SI{8}{GHz}$ and injecting a signal probe at $\omega_\text{s} = 2\pi \cdot \SI{3.8}{GHz}$. Alongside the expected idler peak at $\omega_\text{i} = \omega_\text{p} - \omega_\text{s} = 2\pi \cdot \SI{4.2}{GHz}$, the upconverted sidebands at $\omega_\text{i}+\omega_\text{p} = 2\pi \cdot \SI{11.8}{GHz}$ and $\omega_\text{i}+\omega_\text{p} = 2\omega_\text{p} - \omega_\text{i} = 2\pi \cdot \SI{12.2}{GHz}$ are clearly visible. Although the highest mode lies slightly outside the nominal 4-12 GHz range of the isolators, the effective usable bandwidth was verified prior to characterization. This was done with dedicated statistical analysis of the transmission parameter's standard deviation across the \textit{thru} reference (over 100 repeated measurements), which confirmed sufficient signal integrity and measurement reliability in the 3.6-12.4 GHz range.

\subsection{Flux Line calibration}
The JTWPA employed in the experiment, AI-JTWPA-C by Arctic Instruments \cite{TWPA_patent, JTWPA_patent, VTT_TWPA_paper}, has an on-chip flux line for delivering the flux DC bias needed to tune the nonlinear coefficients of the meta-material. 

This line was experimentally controlled with a DC voltage source. To translate between generated voltage and on-chip induced magnetic flux, the transmission parameter $S_{21}$ of the unpumped device was measured as a function of the generated voltage. The most visible effect is a modulation in its phase, as shown in Fig. \ref{fig:flux_cal}. The two clear dips in the modulations are identified as $\pm0.5\Phi_0$, respectively.

\subsection{Identification of 3-Wave Mixing sweet spot}\label{sec:3WM_sweet_spot}
\begin{figure}[h!]
    \begin{minipage}{0.48\textwidth}
        \centering
        \includegraphics[width=\linewidth]{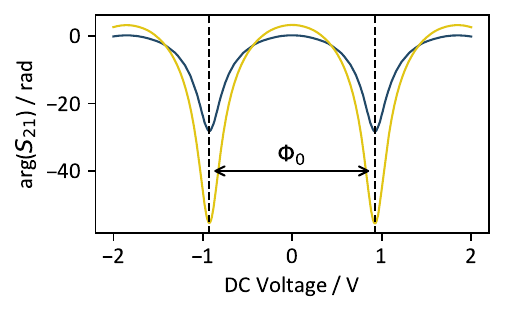}
        \caption{Modulation of the phase of the transmission of the unpumped JTWPA, shown at 4 and 8 GHz as representatives (blue and gold curves, respectively).}
    \label{fig:flux_cal}
    \end{minipage}
    \hfill
    \begin{minipage}{0.48\textwidth}
        \centering
        \includegraphics[width=\linewidth]{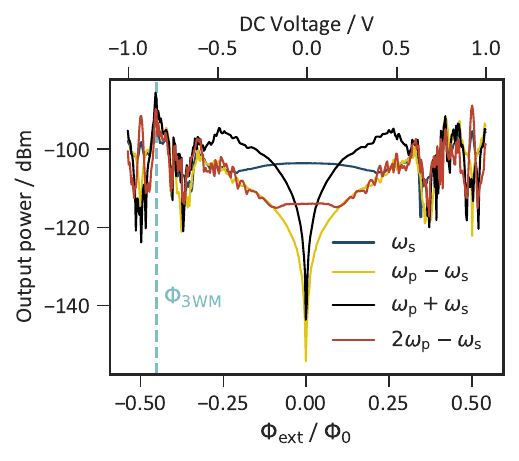}
        \caption{Modulation of 3WM idler as a function of DC bias.  The power levels were scaled to the JTWPA output port with the system gain extracted from the Planck Spectroscopy.
} 
\label{fig:3WM_idler_modulation}
    \end{minipage}
\end{figure}

In order to maximize the 3-Wave Mixing coupling coefficient, the device was pumped at the previously defined pump frequency $\omega_\text{p}=2\pi\cdot$8 GHz, and probed inside the operating band, at $\omega_\text{s}=2\pi\cdot$3.8 GHz. The output at the idler $\omega_\text{i}=\omega_\text{p}-\omega_\text{s}$ 4.2 GHz was measured as a function of the DC bias delivered to the AI-TWPA-C flux line, as shown in Fig. \ref{fig:3WM_idler_modulation}. 

We observe that without flux there is virtually no scattering from the signal to the 3WM idler $\omega_\text{i}$, which 
is instead maximized at $\phi_\text{3WM}=-0.453\Phi_0$, which was used as the working point for the rest of the experiment.

\subsection{Identification of probing power}
\begin{figure}[h!]
\centering
\includegraphics[width=0.85\linewidth]{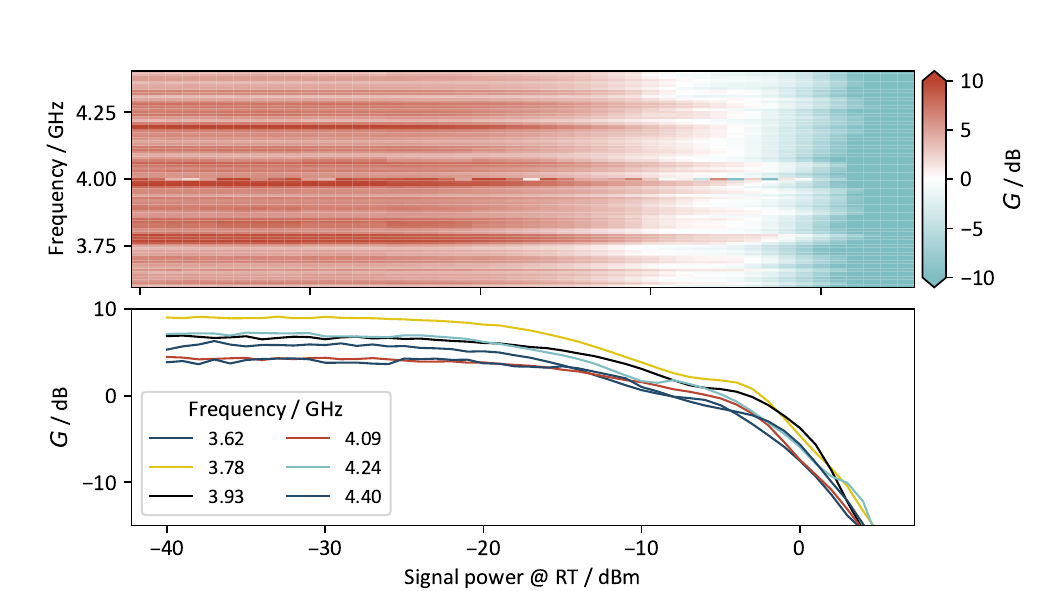}
\caption{Gain as a function of signal probing power delivered by the NVNA. On the top, the data is shown as a colormap, while on the bottom we show gain compression curves for some representative frequencies inside the band of the top plot.
} 
\label{fig:compression}
\end{figure}

In order to ensure that the amplifier is characterized in a linear regime of operation, its gain was measured while varying the signal power level,  see Fig. \ref{fig:compression}. The probing power was chosen to be -40 dBm at the reference plane of the VNA.

\section{VTS for Planck Spectroscopy and readout chain calibration}\label{sec:Planck_spectr}

\begin{figure}[h!]
    \begin{minipage}{0.45\textwidth}
        \centering
        \includegraphics[width=\linewidth]{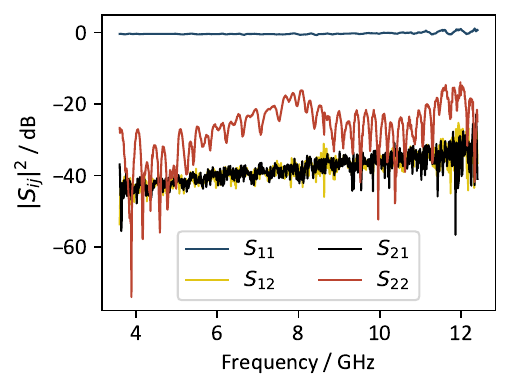}
        \caption{The VTS S-parameters at the switch throws}
        \label{fig:VTS_Sparams}
    \end{minipage}
    \hfill
    \begin{minipage}{0.51\textwidth}
        \centering
        \includegraphics[width=\linewidth]{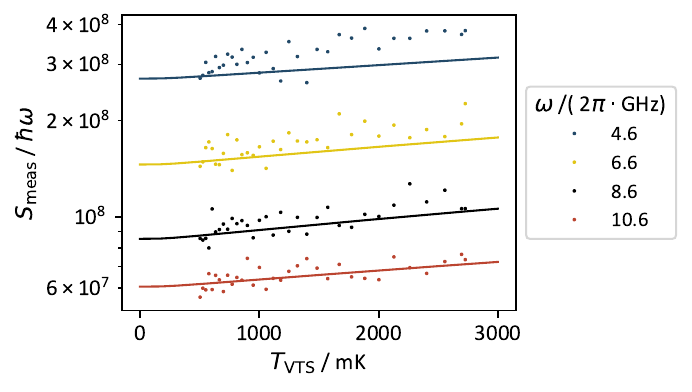}
        \caption{The measured PSD $S_\text{meas}$ as a function of the VTS temperature $T_\text{VTS}$ for some representative frequencies. Dots are data and solid lines are the fits.}
    \label{fig:planck_spectroscopy}
    \end{minipage}
\end{figure}

Step 4 of our experimental workflow is measuring the VTS's S-parameters, shown in Fig. \ref{fig:VTS_Sparams}. Due to a malfunction of the RF switch connected to port 1 of the attenuator, its $S_{11}$ is almost unitary, and $S_{21}$ is well below the expected -20 dB. As argued in the main paper, the requirement to calibrate readout chain $p=1,2$ is 
\begin{equation}\label{eq:VTS_matching}
    |S_{pq}|^2\ll1-|S_{pp}|^2, \quad q \neq p
\end{equation}
This is fulfilled only with $p=2,q=1$, thus only readout chain 2 could be calibrated with our VTS. Furthermore, since $S_{22}$ shows a satisfactory matching on port 2 with very low transmission, the effective behavior of the attenuator during the experiment was even more similar to that of a load.

In order to calibrate the readout chains, the temperature of the VTS was increased with its heater, and the noise PSD was measured with the VNA, employed in Spectrum Analyzer mode, while the VTS slowly cooled down. 

\begin{figure}[h!]
    \centering
    \includegraphics[width=\textwidth]{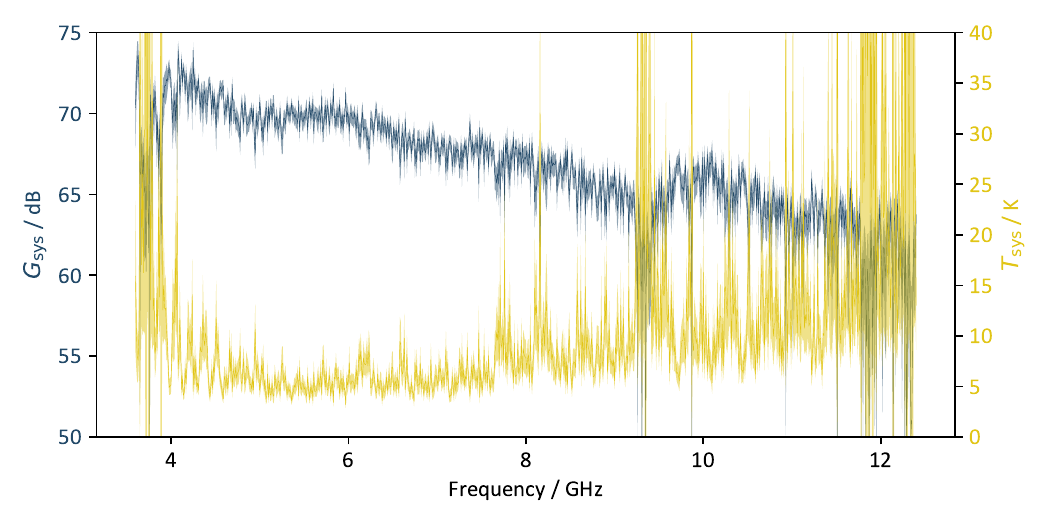}
    \caption{Extrapolated gain and added noise of readout chain 2. The shaded areas represent fit uncertainties.}
    \label{fig:readout_chain_params}
\end{figure}

The power wave $B_2$ was measured with an intermediate frequency bandwidth $B$=20 kHz. The PSD $S_\text{meas}$ was measured while varying the VTS temperature $T_\text{VTS}$, and was fitted to the model:
 
\begin{equation}
    S_\text{meas}(\omega, T_\text{VTS})= G_\text{sys}(\omega)\left[ 
    \Big(1-|S_{22}(\omega)|^2\Big)\frac{\hbar \omega}{2}
    \coth\left(\frac{\hbar \omega}{2 k_B T_\text{VTS}}\right)+\frac{\hbar \omega}{2}|S_{21}(\omega)|^2 + k_B T_\text{sys}(\omega)
    \right]
\end{equation}
as shown in Fig. \ref{fig:planck_spectroscopy}. The results of the fit for all measured frequencies are shown in Fig. \ref{fig:readout_chain_params}.

\section{Synthesis of Measured Photon Flux with Noise Source at the DUT Input}
\label{app:synth}

The theoretical framework and models derived in Appendix B of the main text are used to generate synthetic data corresponding to the serial configuration with the noise source at the DUT input. Such a measurement was not performed in our setup, but all the quantities can be estimated in an informed way either through other measurements or using quantities on the components' datasheets.

We consider a signal at frequency $\omega_\text{s}~=~2~\pi~\cdot~\SI{4}{GHz}$. As previously discussed, the fluctuations on the signal scattered out of the JTWPA are the result of the scattering of input fluctuations at other frequencies. In our case, we identify these spuriously coupled mode with the first-order intermodulation products of signal and pump: $\omega_n = |\omega_\text s+n\omega_\text p|, n\in\mathbb Z$. In addition to the idler, we consider the sidebands  $\omega_\text{p}+\omega_\text{s}, \:2\omega_\text{p}-\omega_\text{s},\:2\omega_\text{p}+\omega_\text{s},\:3\omega_\text{p}+\omega_\text{s},\:4\omega_\text{p}-\omega_\text{s},\:4\omega_\text{p}+\omega_\text{s},\:5\omega_\text{p}-\omega_\text{s}$, which are the upconverted counterparts of the signal and the idler.

The noise at these four modes is generated by a VTS, whose temperature $T$ is sampled at 10 uniformly spaced values between \SI{20}{mK} and \SI{2}{K}. The resulting noise would then travel through some lossy components before entering the JTWPA. As imagined in Fig. 3a of the main paper, when operating a VTS, one would need a directional coupler, in order to provide a pump and a signal to the amplifier, and a switch in order to measure the gain of the JTWPA at its ports. We take the associated insertion loss values $\text{IL}_\text{DC}(\omega)$ and $\text{SW}_\text{DC}(\omega)$ from the respective datasheets of the same components we employ in our setup \cite{dir_coupler_datasheet, switch_datasheet}.

When the fluctuations enter the JTWPA, they are transmitted forward with scattering amplitudes $|S_{2 1}^{(\mathrm{s} n)}|^2$. The device gain $G=|S_{2 1}^{(\mathrm{s} \mathrm{s})}|^2$ is measured thanks to the S-parameter calibration. From Fig. \ref{fig:3WM_idler_modulation} we deduce that the scattering amplitudes from the signal to the other modes are of the same order in magnitude as the gain $G$, at the selected flux sweet spot $\Phi_\text{3WM}$. For generating this synthetic data, we are interested in the reciprocal processes, i.e. the scattering from the other modes to the signal. Assuming reciprocity would mean considering $|S_{2 1}^{(\mathrm{s} n)}|^2~\simeq~G$. This assumption would heavily push towards an enhancement of the nonlinear distortion we aim to quantify, therefore we conservatively choose to scale these scattering amplitudes down as 
\begin{equation}
    |S_{2 1}^{(\mathrm{s} n)}|^2=\begin{cases}
        G^{\,1/(n+1)} \quad & \text{for } n \geq 0 \\
        (G-1)^{1/|n|} \quad & \text{for } n < 0
    \end{cases}
\end{equation}
to account for the dependence of the scattering amplitudes on different powers of the pump amplitude.

Fluctuations may enter on port 2, i.e. back-propagating with respect to the pump. They are expected to travel through the JTWPA without interacting with the pump, and then a portion of them will be reflected at port 1, and then interact with the pump and undergo parametric processes. In order to account for the cumulated effect of reflection and parametric processes, the scattering amplitudes regarding modes entering on port 2 are set to 
$|S_{2 2}^{(\mathrm{s} n)}|^2~=~|S_{2 1}^{(\mathrm{s} n)}|^2/100$.

We use the experimental values $G_\text{sys}=\SI{70.9}{dB}$ and $T_\text{sys}=\SI{4.65}{K}$, experimentally obtained at 4 GHz by calibrating the readout chain, see Section \ref{sec:Planck_spectr}.

With all these quantities in place, it is possible to generate a set of synthetic data points for the noise PSD that could be measured at RT, enabling comparison between the actual physical behavior and the distortion introduced by forcing linearity to model a readout chain that includes a component whose nonlinearity is unknown--at best--or known to be significant--at worst, as in the case of parametric amplifiers.

\bibliographystyle{ieeetr}
\bibliography{references}

%% file: references.bib
@misc{SupplMat,
  author       = {Celotto, Andrea and Alocco, Alessando and Galvano, Bernardo and Fasolo, Luca and Palumbo, Emanuele and Callegaro, Luca and Oberto, Luca and Livreri, Patrizia and Enrico, Emanuele},
  title        = {Supplementary Materials},
  year         = {2026},
month = {May}
}

@article{clerk,
  title = {Introduction to quantum noise, measurement, and amplification},
  author = {Clerk, A. A. and Devoret, M. H. and Girvin, S. M. and Marquardt, Florian and Schoelkopf, R. J.},
  journal = {Rev. Mod. Phys.},
  volume = {82},
  issue = {2},
  pages = {1155--1208},
  numpages = {0},
  year = {2010},
  month = {Apr},
}

@article{Malnou2021,
  title = {Three-Wave Mixing Kinetic Inductance Traveling-Wave Amplifier with Near-Quantum-Limited Noise Performance},
  author = {Malnou, M. and Vissers, M.R. and Wheeler, J.D. and Aumentado, J. and Hubmayr, J. and Ullom, J.N. and Gao, J.},
  journal = {PRX Quantum},
  volume = {2},
  issue = {1},
  pages = {010302},
  numpages = {17},
  year = {2021},
  month = {Jan},
}

@article{Malnou2024,
    author = {Malnou, M. and Larson, T. F. Q. and Teufel, J. D. and Lecocq, F. and Aumentado, J.},
    title = {Low-noise cryogenic microwave amplifier characterization with a calibrated noise source},
    journal = {Review of Scientific Instruments},
    volume = {95},
    number = {3},
    pages = {034703},
    year = {2024},
    month = {03}
}

@book{X_params_book, 
place={Cambridge}, series={The Cambridge RF and Microwave Engineering Series}, title={X-Parameters: Characterization, Modeling, and Design of Nonlinear RF and Microwave Components}, publisher={Cambridge University Press}, author={Root, David E. and Verspecht, Jan and Horn, Jason and Marcu, Mihai}, year={2013}, collection={The Cambridge RF and Microwave Engineering Series}}

@article{Floquet_TWPA,
  title = {Floquet-Mode Traveling-Wave Parametric Amplifiers},
  author = {Peng, Kaidong and Naghiloo, Mahdi and Wang, Jennifer and Cunningham, Gregory D. and Ye, Yufeng and O'Brien, Kevin P.},
  journal = {PRX Quantum},
  volume = {3},
  issue = {2},
  pages = {020306},
  numpages = {20},
  year = {2022},
  month = {Apr},
  publisher = {American Physical Society},
  doi = {10.1103/PRXQuantum.3.020306},
}

@misc{JTWPA_patent,
  author       = {Visa Vesterinen and Juha Hassel},
  title        = {Josephson traveling wave parametric amplifier},
  number       = {FI20195045A1},
  type         = {Patent},
  date         = {2020-07-25},
  url          = {https://patents.google.com/patent/FI20195045A1}
}

@misc{TWPA_patent,
  author       = {Visa Vesterinen and Slawomir Simbierowicz},
  title        = {Traveling Wave Parametric Amplifier},
  year         = {2021},
  type         = {Patent},
  howpublished = {FI Patent FI20205401A1},
  url          = {https://patents.google.com/patent/FI20205401A1}
}

@article{VTT_TWPA_paper,
  title = {Broadband Continuous-Variable Entanglement Generation Using a Kerr-Free Josephson Metamaterial},
  author = {Perelshtein, M.R. and Petrovnin, K.V. and Vesterinen, V. and Hamedani Raja, S. and Lilja, I. and Will, M. and Savin, A. and Simbierowicz, S. and Jabdaraghi, R.N. and Lehtinen, J.S. and Gr\"onberg, L. and Hassel, J. and Prunnila, M.P. and Govenius, J. and Paraoanu, G.S. and Hakonen, P.J.},
  journal = {Phys. Rev. Appl.},
  volume = {18},
  issue = {2},
  pages = {024063},
  numpages = {14},
  year = {2022},
  month = {Aug},
}

@article{SNAIL,
    author = {Frattini, N. E. and Vool, U. and Shankar, S. and Narla, A. and Sliwa, K. M. and Devoret, M. H.},
    title = {3-wave mixing Josephson dipole element},
    journal = {Applied Physics Letters},
    volume = {110},
    number = {22},
    pages = {222603},
    year = {2017},
    month = {05}
}

@article{Ranzani_calibration,
    author = {Ranzani, Leonardo and Spietz, Lafe and Popovic, Zoya and Aumentado, José},
    title = {Two-port microwave calibration at millikelvin temperatures},
    journal = {Review of Scientific Instruments},
    volume = {84},
    number = {3},
    pages = {034704},
    year = {2013},
    month = {03},
    issn = {0034-6748},
}

@article{100qubits,
  author  = {S. Krinner and S. Storz and P. Kurpiers and P. Magnard and J. Heinsoo and R. Keller and J. L\"utolf and C. Eichler and A. Wallraff},
  title   = {Engineering cryogenic setups for 100-qubit scale superconducting circuit systems},
  journal = {EPJ Quantum Technology},
  year    = {2019},
  volume  = {6},
  number  = {1},
  pages   = {2},
}

@ARTICLE{NPLcal,
  author={Stanley, Manoj and De Graaf, Sebastian and Hönigl-Decrinis, Teresa and Lindström, Tobias and Ridler, Nick M.},
  journal={IEEE Access}, 
  title={Characterizing Scattering Parameters of Superconducting Quantum Integrated Circuits at Milli-Kelvin Temperatures}, 
  year={2022},
  volume={10},
  number={},
  pages={43376-43386},
}

@misc{oberto2026,
      title={Full Two-Port S-Parameters at mK Temperatures: a Calibration Strategy and Uncertainty Budget}, 
      author={Luca Oberto and Ehsan Shokrolahzade and Emanuele Enrico and Luca Fasolo and Andrea Celotto and Bernardo Galvano and Alessandro Alocco and Paolo Terzi and Faisal A. Mubarak and Marco Spirito},
      year={2026},
      eprint={2505.19922},
      archivePrefix={arXiv},
      primaryClass={physics.ins-det},
      url={https://arxiv.org/abs/2505.19922}, 
}

@article{Caves_Schumaker,
  title = {New formalism for two-photon quantum optics. I. Quadrature phases and squeezed states},
  author = {Caves, Carlton M. and Schumaker, Bonny L.},
  journal = {Phys. Rev. A},
  volume = {31},
  issue = {5},
  pages = {3068--3092},
  numpages = {0},
  year = {1985},
  month = {May},
  publisher = {American Physical Society},
  doi = {10.1103/PhysRevA.31.3068},
}

@ARTICLE{Ferrero1992,
  author={Ferrero, A. and Pisani, U.},
  journal={IEEE Microwave and Guided Wave Letters}, 
  title={Two-port network analyzer calibration using an unknown 'thru'}, 
  year={1992},
  volume={2},
  number={12},
  pages={505-507},
  doi={10.1109/75.173410}}

@ARTICLE{Aumentado_review,
  author={Aumentado, Jose},
  journal={IEEE Microwave Magazine}, 
  title={Superconducting Parametric Amplifiers: The State of the Art in Josephson Parametric Amplifiers}, 
  year={2020},
  volume={21},
  number={8},
  pages={45-59},
  doi={10.1109/MMM.2020.2993476}}

@INPROCEEDINGS{Peng_Xparameters,
  author={Peng, Kaidong and Poore, Rick and Krantz, Philip and Root, David E. and O’Brien, Kevin P.},
  booktitle={2022 IEEE International Conference on Quantum Computing and Engineering (QCE)}, 
  title={X-parameter based design and simulation of Josephson traveling-wave parametric amplifiers for quantum computing applications}, 
  year={2022},
  volume={},
  number={},
  pages={331-340},
  doi={10.1109/QCE53715.2022.00054}}

@article{PTB_TWPA,
  title = {rf-SQUID-based traveling-wave parametric amplifier with input saturation power of $\ensuremath{-}84$ dBm across more than one octave in bandwidth},
  author = {Gaydamachenko, Victor and Kissling, Christoph and Gr\"unhaupt, Lukas},
  journal = {Phys. Rev. Appl.},
  volume = {23},
  issue = {6},
  pages = {064053},
  numpages = {18},
  year = {2025},
  month = {Jun},
  publisher = {American Physical Society},
  doi = {10.1103/1qk4-fzkq},
}

@article{Zobrist,
    author = {Zobrist, Nicholas and Eom, Byeong Ho and Day, Peter and Mazin, Benjamin A. and Meeker, Seth R. and Bumble, Bruce and LeDuc, Henry G. and Coiffard, Grégoire and Szypryt, Paul and Fruitwala, Neelay and Lipartito, Isabel and Bockstiegel, Clint},
    title = {Wide-band parametric amplifier readout and resolution of optical microwave kinetic inductance detectors},
    journal = {Applied Physics Letters},
    volume = {115},
    number = {4},
    pages = {042601},
    year = {2019},
    month = {07},
    issn = {0003-6951},
    doi = {10.1063/1.5098469},
}

@article{Martinis2015,
    author = {White, T. C. and Mutus, J. Y. and Hoi, I.-C. and Barends, R. and Campbell, B. and Chen, Yu and Chen, Z. and Chiaro, B. and Dunsworth, A. and Jeffrey, E. and Kelly, J. and Megrant, A. and Neill, C. and O'Malley, P. J. J. and Roushan, P. and Sank, D. and Vainsencher, A. and Wenner, J. and Chaudhuri, S. and Gao, J. and Martinis, John M.},
    title = {Traveling wave parametric amplifier with Josephson junctions using minimal resonator phase matching},
    journal = {Applied Physics Letters},
    volume = {106},
    number = {24},
    pages = {242601},
    year = {2015},
    month = {06},
    issn = {0003-6951},
    doi = {10.1063/1.4922348},
}

@article{HoEom2012,
  author  = {Ho Eom, Byeong and Day, Peter K. and LeDuc, Henry G. and Zmuidzinas, Jonas},
  title   = {A wideband, low-noise superconducting amplifier with high dynamic range},
  journal = {Nature Physics},
  year    = {2012},
  month   = {Aug},
  volume  = {8},
  number  = {8},
  pages   = {623--627},
  doi     = {10.1038/nphys2356},
}

@article{RanzaniKITWPA,
    author = {Ranzani, L. and Bal, M. and Fong, Kin Chung and Ribeill, G. and Wu, X. and Long, J. and Ku, H.-S. and Erickson, R. P. and Pappas, D. and Ohki, T. A.},
    title = {Kinetic inductance traveling-wave amplifiers for multiplexed qubit readout},
    journal = {Applied Physics Letters},
    volume = {113},
    number = {24},
    pages = {242602},
    year = {2018},
    month = {12},
    issn = {0003-6951},
    doi = {10.1063/1.5063252},
}

@misc{klimovich2023,
      title={Demonstration of a Quantum Noise Limited Traveling-Wave Parametric Amplifier}, 
      author={Nikita Klimovich and Peter Day and Shibo Shu and Byeong Ho Eom and Henry Leduc and Andrew Beyer},
      year={2023},
      eprint={2306.11028},
      archivePrefix={arXiv},
      primaryClass={quant-ph},
      url={https://arxiv.org/abs/2306.11028}, 
}

@INPROCEEDINGS{RF_transfer,
  author={Celep, Murat and Shin, Sang-Hee and Stanley, Manoj and Breakenridge, Eric and Singh, Suren and Ridler, Nick},
  booktitle={2024 Conference on Precision Electromagnetic Measurements (CPEM)}, 
  title={SI Traceable RF and Microwave Power Measurements at Cryogenic Temperatures}, 
  year={2024},
  volume={},
  number={},
  pages={1-2},
  doi={10.1109/CPEM61406.2024.10646150}}

@article{Royal_Holloway,
  title = {Two-Level System as a Quantum Sensor for Absolute Calibration of Power},
  author = {H\"onigl-Decrinis, T. and Shaikhaidarov, R. and de Graaf, S.E. and Antonov, V.N. and Astafiev, O.V.},
  journal = {Phys. Rev. Appl.},
  volume = {13},
  issue = {2},
  pages = {024066},
  numpages = {9},
  year = {2020},
  month = {Feb},
  publisher = {American Physical Society},
  doi = {10.1103/PhysRevApplied.13.024066}
}

@article{2D_Planck,
  title = {Two-dimensional Planck spectroscopy for microwave photon calibration},
  author = {Gandorfer, S. and Renger, M. and Yam, W.K. and Fesquet, F. and Marx, A. and Gross, R. and Fedorov, K.G.},
  journal = {Phys. Rev. Appl.},
  volume = {23},
  issue = {2},
  pages = {024064},
  numpages = {10},
  year = {2025},
  month = {Feb},
  publisher = {American Physical Society},
  doi = {10.1103/PhysRevApplied.23.024064},
  url = {https://link.aps.org/doi/10.1103/PhysRevApplied.23.024064}
}

@article{SNTJ_Spietz,
    author = {Spietz, Lafe and Schoelkopf, R. J. and Pari, Patrick},
    title = {Shot noise thermometry down to 10mK},
    journal = {Applied Physics Letters},
    volume = {89},
    number = {18},
    pages = {183123},
    year = {2006},
    month = {11},
    issn = {0003-6951},
    doi = {10.1063/1.2382736},
}

@INPROCEEDINGS{SNTJ_Aumentado,
  author={Su-Wei Chang and Aumentado, Jose and Wei-Ting Wong and Bardin, Joseph C.},
  booktitle={2016 IEEE MTT-S International Microwave Symposium (IMS)}, 
  title={Noise measurement of cryogenic low noise amplifiers using a tunnel-junction shot-noise source}, 
  year={2016},
  volume={},
  number={},
  pages={1-4},
  doi={10.1109/MWSYM.2016.7538226}}

@article{Ranadive2022,
  author  = {Ranadive, Arpit and Esposito, Martina and Planat, Luca and Bonet, Edgar and Naud, Cécile and Buisson, Olivier and Guichard, Wiebke and Roch, Nicolas},
  title   = {Kerr reversal in Josephson meta-material and traveling wave parametric amplification},
  journal = {Nature Communications},
  year    = {2022},
  month   = {Apr},
  volume  = {13},
  number  = {1},
  pages   = {1737},
  doi     = {10.1038/s41467-022-29375-5},
  url     = {}
}

@article{Planat,
  title = {Photonic-Crystal Josephson Traveling-Wave Parametric Amplifier},
  author = {Planat, Luca and Ranadive, Arpit and Dassonneville, R\'emy and Puertas Mart\'{\i}nez, Javier and L\'eger, S\'ebastien and Naud, C\'ecile and Buisson, Olivier and Hasch-Guichard, Wiebke and Basko, Denis M. and Roch, Nicolas},
  journal = {Phys. Rev. X},
  volume = {10},
  issue = {2},
  pages = {021021},
  numpages = {19},
  year = {2020},
  month = {Apr},
  publisher = {American Physical Society},
  doi = {10.1103/PhysRevX.10.021021},
  url = {https://link.aps.org/doi/10.1103/PhysRevX.10.021021}
}

@manual{switch_datasheet,
    title        = {Cryogenic SP6T Subminiature SMA 18GHz Latching 12Vdc Pins Terminals Bipolar Actuator Command, Part Number R5927B2141},
    author       = {Radiall},
    year         = {2022},
    month        = {October},
    organization = {Radiall},
    url          = {https://www.radiall.com/cryogenic-sp6t-subminiature-sma-18ghz-latching-12vdc-pins-terminals-bipolar-actuator-command-r5927b2141.html}
}

@manual{dir_coupler_datasheet,
    title        = {{
QMC-CRYOCOUPLER-10 - Cryogenic Directional Coupler 10 dB | 4 GHz to 12 GHz | Gold Plated}},
    author       = {Quantum\hphantom{ }Microwave},
    organization = {Quantum Microwave},
    url          = {https://quantummicrowave.com/product/4-ghz-to-12-ghz-cryogenic-directional-coupler/}
}

@article{ranadive2025,
author = {Ranadive, Arpit and Fazliji, Bekim and Gal, Gwenael and Cappelli, Giulio and Butseraen, Guilliam and Bonet, Edgar and Eyraud, Eric and Böhling, Sina and Planat, Luca and Metelmann, A. and Roch, Nicolas},
title = {A travelling-wave parametric amplifier isolator},
journal = {Nature Electronics},
year = {2025},
doi = {10.1038/s41928-025-01489-w},
url = {http://dx.doi.org/10.1038/s41928-025-01489-w}
}

@article{malnou2025,
author = {Malnou, M. and Miller, B. T. and Estrada, J. A. and Genter, K. and Cicak, K. and Teufel, J. D. and Aumentado, J. and Lecocq, F.},
title = {A travelling-wave parametric amplifier and converter},
journal = {Nature Electronics},
year = {2025},
doi = {10.1038/s41928-025-01445-8},
url = {http://dx.doi.org/10.1038/s41928-025-01445-8}
}

@article{Chapman2017,
  title = {Widely Tunable On-Chip Microwave Circulator for Superconducting Quantum Circuits},
  author = {Chapman, Benjamin J. and Rosenthal, Eric I. and Kerckhoff, Joseph and Moores, Bradley A. and Vale, Leila R. and Mates, J. A. B. and Hilton, Gene C. and Lalumi\`ere, Kevin and Blais, Alexandre and Lehnert, K. W.},
  journal = {Phys. Rev. X},
  volume = {7},
  issue = {4},
  pages = {041043},
  numpages = {16},
  year = {2017},
  month = {Nov},
  publisher = {American Physical Society},
  doi = {10.1103/PhysRevX.7.041043},
  url = {https://link.aps.org/doi/10.1103/PhysRevX.7.041043}
}

@article{Lecocq2017,
  title = {Nonreciprocal Microwave Signal Processing with a Field-Programmable Josephson Amplifier},
  author = {Lecocq, F. and Ranzani, L. and Peterson, G. A. and Cicak, K. and Simmonds, R. W. and Teufel, J. D. and Aumentado, J.},
  journal = {Phys. Rev. Appl.},
  volume = {7},
  issue = {2},
  pages = {024028},
  numpages = {17},
  year = {2017},
  month = {Feb},
  publisher = {American Physical Society},
  doi = {10.1103/PhysRevApplied.7.024028},
  url = {https://link.aps.org/doi/10.1103/PhysRevApplied.7.024028}
}

@article{Sliwa2015,
  title = {Reconfigurable Josephson Circulator/Directional Amplifier},
  author = {Sliwa, K. M. and Hatridge, M. and Narla, A. and Shankar, S. and Frunzio, L. and Schoelkopf, R. J. and Devoret, M. H.},
  journal = {Phys. Rev. X},
  volume = {5},
  issue = {4},
  pages = {041020},
  numpages = {10},
  year = {2015},
  month = {Nov},
  publisher = {American Physical Society},
  doi = {10.1103/PhysRevX.5.041020},
  url = {https://link.aps.org/doi/10.1103/PhysRevX.5.041020}
}

@article{Demarets2026,
  title = {Broadband magnetless isolation in a flux-pumped, dispersion-engineered transmission line},
  author = {Demarets, M. and Vadiraj, A. M. and Caloz, C. and Greve, K. De},
  journal = {Phys. Rev. Appl.},
  pages = {},
  year = {2026},
  month = {May},
  publisher = {American Physical Society},
  doi = {10.1103/vjy5-wxpy},
  url = {https://link.aps.org/doi/10.1103/vjy5-wxpy}
}
